# Designing a K-state P-bit Engine

Mohammad Khairul Bashar, Abir Hasan, Nikhil Shukla*

Department of Electrical and Computer Engineering, University of Virginia,

Charlottesville, VA- 22904 USA

*e-mail: ns6pf@virginia.edu

**ABSTRACT**

**Probabilistic bit (p-bit)-based compute engines utilize the unique capability of a p-bit to probabilistically switch between two states to solve computationally challenging problems. However, when solving problems that require more than two states (e.g., problems such as Max-3-Cut, verifying if a graph is K-partite (K>2) etc.), additional pre-processing steps such as graph reduction are required to make the problem compatible with a two-state p-bit platform. Moreover, this not only increases the problem size by entailing the use of auxiliary variables but can also degrade the solution quality. In this work, we develop a unique framework for implementing a K-state (K>2) p-bit engine. Furthermore, from an implementation standpoint, we show that such a K-state p-bit engine can be implemented using N traditional (2-state) p-bits, and one multi-state p-bit- a novel concept proposed here. Augmenting traditional p-bit platforms, our approach enables us to solve an archetypal combinatoric problem class requiring multiple states, namely Max-K-Cut (K=3, 4 shown here), without using any additional auxiliary variables. Thus, our work fundamentally advances the functional capability of p-bit engines, enabling them to solve a broader class of computationally challenging problems more efficiently.**

## I. INTRODUCTION

Modern information processing has relied largely on digital computers based on the von Neumann architecture. However, with the explosion in data and its applications, there is an increasing demand for solving computational problems, many of which are still considered challenging in the digital computing framework. This top-down shift driven by applications, combined with the slowing down of Moore's law at the hardware end, has created a strong impetus to explore non von Neumann computing paradigms that can complement digital computing.

The search for alternate computational approaches has led to a plethora of models and corresponding hardware platforms that range from quantum annealers [1], optical devices (e.g., DOPO: degenerate optical parametric oscillators) [2], electronic devices such as synchronized oscillators [3]-[4], coupled latches [5] to biologically inspired computation [6], each with their distinctive features. Amongst them, probabilistic computing [7], based on probabilistic bits (p-bits), has garnered significant attention owing to its versatility, and compatibility with CMOS as well as emerging technology platforms such as magnetic tunnel junctions (MTJs). A manifestation of stochastic binary neural networks, p-bit engines have been used to accelerate a variety of applications ranging from deep Boltzmann networks [8] to computationally intractable problems in combinatorial optimization (e.g., integer factorization [9], minimization of the Ising Hamiltonian [10]).

The p-bit essentially denotes a bit that probabilistically switches between *two states*, typically represented by '-1' / '1', or '0' / '1'. From a computational standpoint, the number of states (two here) signifies the representational capability of a p-bit, an important consideration when mapping a problem to a p-bit based computing platform. As a case in point, the *two sets* created when solving the MaxCut of a graph can be elegantly mapped to the *two states* of the p-bit; computing the MaxCut of a graph is an archetypal combinatorial optimization problem that entails dividing the nodes of a graph into two sets such that the total weight of the edges common to both the sets is maximized. The same is true for other problems such as integer factorization where the binary digits (0 and 1) can also be directly mapped to the two states of the p-bit. However, such a mapping no longer remains straightforward when the problem requires more than two states. For instance, computing the Max-3-Cut, an extension of the MaxCut problem, entails the creation of three sets such that the number of edges common to any two sets is maximized [11]. Along similar lines, verifying if a graph is tripartite (three colorable) entails the use of *three states*, which are not directly available in the traditional p-bit. Such problems find extensive practical use in areas such as resource allocation [12].

Overcoming this constraint then entails additional processing steps such as using graph reduction techniques that reduce the problem to a format that is compatible with the two-state property of the p-bits [13]. We note that other computational platforms such as qubit-based quantum annealers as well as oscillator Ising machines also exhibit the same characteristics [14], [11]. Such added pre-processing entails the introduction of auxiliary variables (that need additional p-bits) which effectively increases the size of the problem that must be solved by the p-bit platform. For instance, solving the Max-3-Cut problem on a graph of N nodes will typically entail using up to 3N p-bits [15]. This not only results in increased computational resources but also degrades the solution quality. To circumvent this challenge, we first develop the theoretical framework for a K-state p-bit engine and showcase its ability to directly solve the Max-K-Cut problem (K=3,4) – a prototypical combinatorial optimization problem that uses K-states, without graph reduction. Subsequently, to implement such a K-state p-bit engine, we propose a novel multi-state p-bit, and illustrate a scheme to realize it by exploiting the stochastic threshold switching across the insulator-metal transition (IMT) in $VO_2$.

## II. RESULTS

**Theoretical framework of a K-state (K>2) p-bit engine:** We first illustrate the formulation of a K-state p-bit engine using K=3, and subsequently, generalize it to K states.

To develop a 3-state p-bit engine, we first consider a p-bit ($\bar{\sigma}$) with three states, wherein the three states are defined using three orthogonal unit vectors ($\widehat{\boldsymbol{e_1}}, \widehat{\boldsymbol{e_2}}, \widehat{\boldsymbol{e_3}}$). The p-bit is designed such that if and when a bit transitions from its existing state, it does so to one of the other M (= K-1; here, two) states with equal probability. Using these characteristics, the objective function for computing the Max-3-Cut can now be expressed by the following energy function:

$$H = -\sum_{i<j} J_{ij}\left(2\overline{\sigma_i}.\overline{\sigma_j} - 1\right) \tag{1}$$

where $[J]$ is the adjacency matrix for the input graph. Solving the Max-3-Cut problem entails anti-ferromagnetic coupling i.e., $J_{ij} = -1(0)$ when an edge is present (absent) between nodes $i$ and $j$, respectively; we consider unweighted graphs in this work. Equation (1) shows that when the nodes containing an edge lie in the same set (i.e., the two corresponding p-bits exhibit the same state), $\overline{\sigma_i}.\overline{\sigma_i} = 1$, which then increases the energy (by one unit). In contrast, when the edge lies in two different states (i.e., the two corresponding p-bits are in different states) $\overline{\sigma_i}.\overline{\sigma_j} = 0\ (i \neq j)$ which subsequently, decreases the energy (by one unit). Thus, solving the Max-3-Cut solution is equivalent to computing the ground state (minima) of equation (1).

Using an approach similar to the conventional p-bit engines, the weighted synaptic feedback ($\phi$) to the $\alpha^{th}$ p-bit can be formulated as $\phi_\alpha = \sum_{j=1, j\neq\alpha}^{N} J_{\alpha j}\left(2\overline{\sigma_\alpha}.\overline{\sigma_j} - 1\right)$. The resulting output of the 3-state p-bit is expressed as:

$$\overline{\sigma^{+}_{\alpha,3-\mathrm{state}}} =$$

$$(\overline{\sigma_\alpha}.\widehat{\boldsymbol{e_1}})\left[\left[0.5\left(1+\mathrm{sgn}(f(\phi_\alpha))\right)\right]\widehat{\boldsymbol{e_1}} + \left[0.5\left(1-\mathrm{sgn}(f(\phi_\alpha))\right)g\right]\widehat{\boldsymbol{e_2}} + \left[0.5\left(1-\mathrm{sgn}(f(\phi_\alpha))\right)(1-g)\right]\widehat{\boldsymbol{e_3}}\right]$$

$$+$$

$$(\overline{\sigma_\alpha}.\widehat{\boldsymbol{e_2}})\left[\left[0.5\left(1+\mathrm{sgn}(f(\phi_\alpha))\right)\right]\widehat{\boldsymbol{e_2}} + \left[0.5\left(1-\mathrm{sgn}(f(\phi_\alpha))\right)g\right]\widehat{\boldsymbol{e_3}} + \left[0.5\left(1-\mathrm{sgn}(f(\phi_\alpha))\right)(1-g)\right]\widehat{\boldsymbol{e_1}}\right] \tag{2a}$$

$$+$$

$$(\overline{\sigma_\alpha}.\widehat{\boldsymbol{e_3}})\left[\left[0.5\left(1+\mathrm{sgn}(f(\phi_\alpha))\right)\right]\widehat{\boldsymbol{e_3}} + \left[0.5\left(1-\mathrm{sgn}(f(\phi_\alpha))\right)g\right]\widehat{\boldsymbol{e_1}} + \left[0.5\left(1-\mathrm{sgn}(f(\phi_\alpha))\right)(1-g)\right]\widehat{\boldsymbol{e_2}}\right]$$

Where,

$$f(\phi_\alpha) = \tanh\left(\beta(\phi_\alpha)\right) - \vartheta_{[-1,1]} \quad (2b)$$

$f(\phi_\alpha)$ represents the non-linear transformation on the synaptic input $\phi_\alpha$ along with a stochastic component represented by $\vartheta_{[-1,1]}$ which represents a random number taken from a uniform distribution from -1 to 1. $g = r_{\{0,1\}}$, where $r_{\{0,1\}}$ denotes a random selection between the two values, 0 and 1, with equal probability. $\beta$ in the $\tanh()$ function essentially controls the steepness of the non-linear transformation and can be considered as representing the 'inverse temperature' in the Boltzmann statistics.

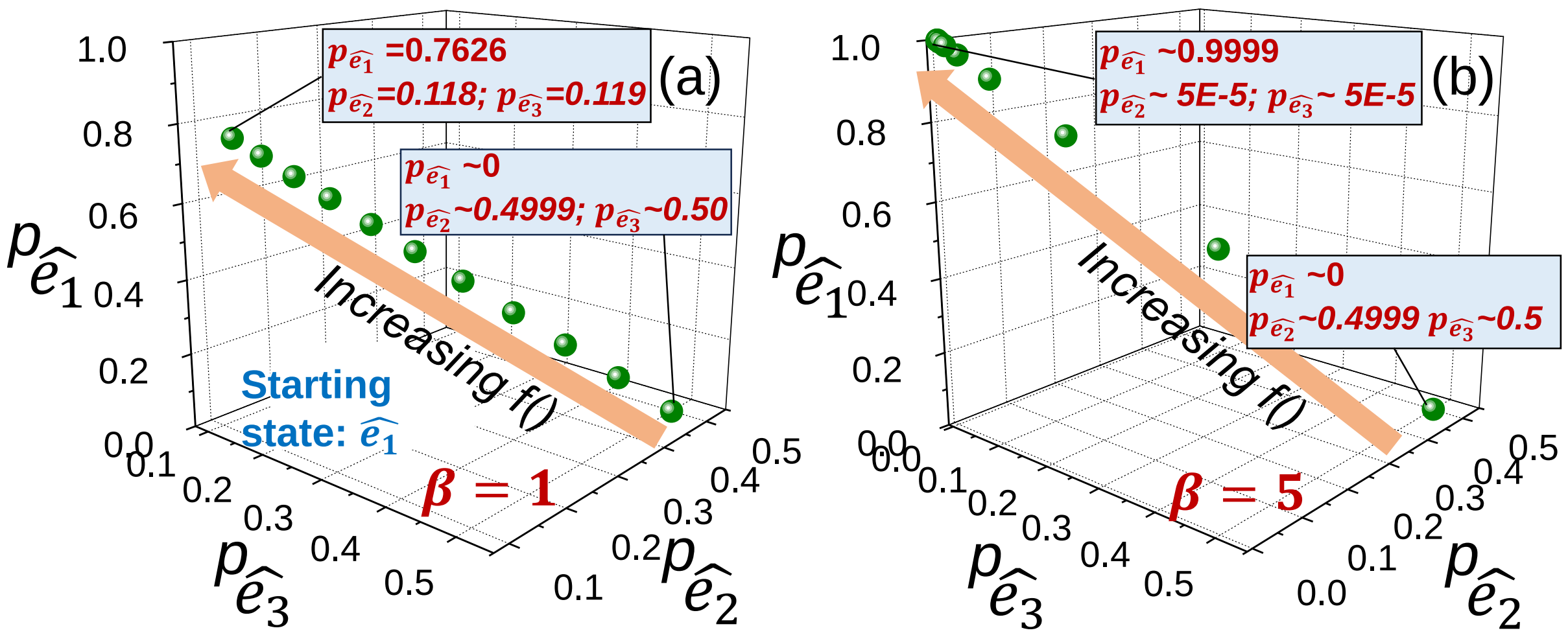

**Fig. 1**: Evolution of the probability of the three states $(\widehat{e_1}, \widehat{e_2}, \widehat{e_3})$ of a 3-state p-bit as a function of $f()$ for (a) $\beta = 1$ (b) $\beta = 5$. The starting state is assumed to be $\widehat{e_1}$ and the probabilities are calculated using $10^6$ trials.

The behavior of the K-state p-bit described in equation (2a) can be understood as follows. If the p-bit is presently in a particular state, say $\widehat{e_1}$, then the second and third terms on the right-hand side (RHS) of the equation (2a) evaluate to zero since $(\overline{\sigma_\alpha}.\widehat{e_2}) = (\overline{\sigma_\alpha}.\widehat{e_3}) = 0$. Similarly, if the current state of the p-bit is $\widehat{e_2}$ $(\widehat{e_3})$, then the first and third (first and second) terms on the RHS of equation (2a) evaluate to 0, respectively. Furthermore, the synaptic input $\phi$ and $f()$ are designed such that when the present state is 'favored', $f()$ tends to 1, whereas when a switch to a different state (here, $\widehat{e_2}$ $or$ $\widehat{e_3}$) is favorable, $f(\,)$ tends to 0. Moreover, the transition to the other M (=K-1, here 2) states is designed to be equally probable since $g$ $(= r_{\{0,1\}})$ is drawn from a discrete uniform distribution. Figures 1a,b show the simulated probability of a particular state as a function of the synaptic input for $\beta$ = 1 and $\beta$ = 5, respectively, considering the initial state to be $\widehat{e_1}$. It can be observed that not only does the probability of switching to a different state increase as the synaptic input decreases (by design), but the probability of switching to any one of the other (two) states is also nearly equal. The probability for a given state is calculated using $10^6$ trials.

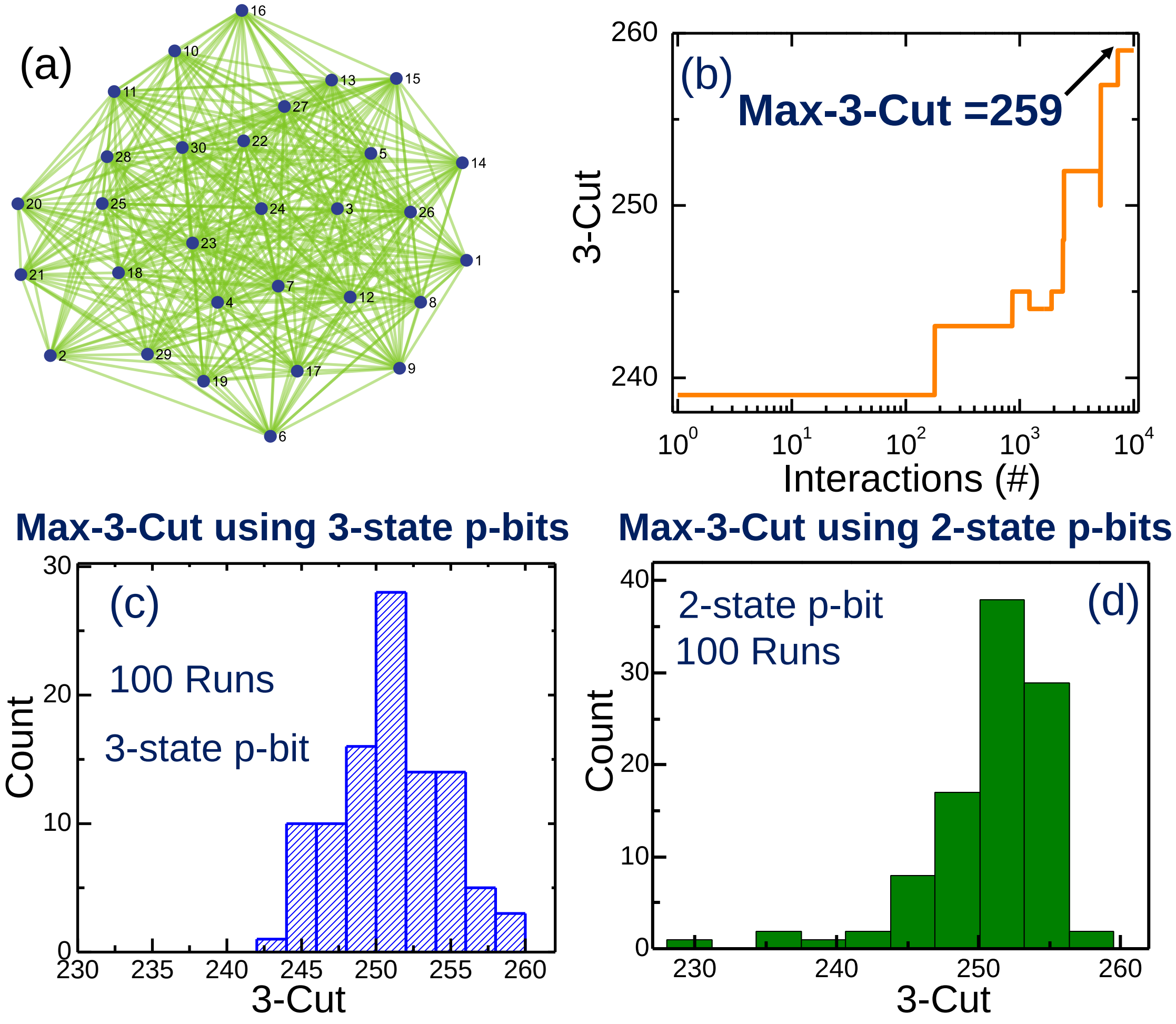

**Fig. 2:** (a) A 30 node representative graph. (b) Evolution of the 3-Cut with the number of iterations, obtained using the proposed 3-state p-bit based solver. Distribution of Max-3-Cut computed using the (c) 3-state p-bit, and (d) 2-state p-bit over 100 trials. For the 2-state p-bit solver, the input graph is reduced to a larger graph with 90 nodes, whereas the problem can be directly solved using the 3-state p-bits. (β=1 was used in the 3-state p-bit and 2-state p-bit simulations).

Using the approach developed above, we evaluate the Max-3-Cut for an illustrative graph with 30 nodes shown in Fig. 2a. Fig 2b shows the calculated 3-cut for a representative trial while Fig. 2c shows the distribution of the Max-3-Cut solution obtained over 100 trials. We also compare our solution with that obtained by using the traditional approach that entails decomposing the graph to a form compatible with the Ising formulation, which increases the number of nodes to 90. Subsequently, this is followed by using the traditional 2-state p-bit framework to compute the solution (Fig. 2d). Comparing Fig. 2c and Fig. 2d, it can be observed that the proposed method enables computation of the Max-3-Cut with a smaller number of nodes while yielding comparable solution quality.

Next, we extend the theoretical framework developed above for a 3-state p-bit engine to a probabilistic engine with K states. The state update rule is then expressed as:

$$
\begin{aligned}
&\overline{\sigma^{+}_{\alpha,\mathrm{K-state}}} = \\
&(\overline{\sigma_{\alpha}}.\widehat{\boldsymbol{e_1}})\left[\left\{\left[0.5\left(1+\mathrm{sgn}\left(f(\phi_\alpha)\right)\right)\right]\widehat{\boldsymbol{e_1}}\right\}+\left[0.5\left(1-\mathrm{sgn}\left(f(\phi_\alpha)\right)\right)g_1\right]\widehat{\boldsymbol{e_2}}+\left[0.5\left(1-\mathrm{sgn}\left(f(\phi_\alpha)\right)\right)\ g_2\right]\widehat{\boldsymbol{e_3}}+\cdots\right.\\
&\qquad\qquad\left.+\left[0.5\left(1-\mathrm{sgn}\left(f(\phi_\alpha)\right)\right)\ g_{K-1}\right]\widehat{\boldsymbol{e_K}}\right]\\
&\qquad\qquad\qquad\qquad+\\
&(\overline{\sigma_{\alpha}}.\widehat{\boldsymbol{e_2}})\left[\left\{\left[0.5\left(1+\mathrm{sgn}\left(f(\phi_\alpha)\right)\right)\right]\widehat{\boldsymbol{e_2}}\right\}+\left\{\left[0.5\left(1-\mathrm{sgn}\left(f(\phi_\alpha)\right)\right)g_1\right]\widehat{\boldsymbol{e_1}}\right\}+\left[0.5\left(1-\mathrm{sgn}\left(f(\phi_\alpha)\right)\right)\ g_2\right]\widehat{\boldsymbol{e_3}}+\cdots\right.\\
&\qquad\qquad\left.+\left[0.5\left(1-\mathrm{sgn}\left(f(\phi_\alpha)\right)\right)\ g_{K-1}\right]\widehat{\boldsymbol{e_K}}\right]\\
&\qquad\qquad\qquad\qquad+ \qquad\qquad\qquad\qquad\qquad\qquad\qquad\qquad \text{(3a)}\\
&\qquad\qquad\qquad\qquad\ldots\\
&\qquad\qquad\qquad\qquad+\\
&(\overline{\sigma_{\alpha}}.\widehat{\boldsymbol{e_K}})\left[\left\{\left[0.5\left(1+\mathrm{sgn}\left(f(\phi_\alpha)\right)\right)\right]\widehat{\boldsymbol{e_K}}\right\}+\left[0.5\left(1-\mathrm{sgn}\left(f(\phi_\alpha)\right)\right)g_1\right]\widehat{\boldsymbol{e_1}}+\left[0.5\left(1-\mathrm{sgn}\left(f(\phi_\alpha)\right)\right)\ g_2\right]\widehat{\boldsymbol{e_2}}+\cdots\right.\\
&\qquad\qquad\left.+\left[0.5\left(1-\mathrm{sgn}\left(f(\phi_\alpha)\right)\right)\ g_{K-1}\right]\widehat{\boldsymbol{e_{K-1}}}\right]
\end{aligned}
$$

$$\overline{\sigma^{+}_{\alpha,\mathrm{K-state}}} = \sum_{i=1}^{K}\left[(\overline{\sigma_{\alpha}}.\hat{e_i})\left\{\left[0.5\left(1+\mathrm{sgn}\big(f(\phi_\alpha)\big)\right)\right]\hat{e_i} + \sum_{j=1}^{i-1}\left[\left[0.5\left(1-\mathrm{sgn}\big(f(\phi_\alpha)\big)\right)g_j\right]\hat{e_j}\right] + \sum_{j=i+1}^{K}\left[\left[0.5\left(1-\mathrm{sgn}\big(f(\phi_\alpha)\big)\right)g_{j-1}\right]\hat{e_j}\right]\right\}\right] \quad (3b)$$

where $g = \{g_1, g_2, \dots, g_{K-1}\}$ is a random one hot vector of length M= $(K-1)$, i.e., only one element of $g$ is 1 while the rest are 0. Essentially $g$ is defined such that the state to which the bit will probabilistically transition, is also randomly selected. Here, $f(\phi_\alpha)$ is the non-linear transform on the synaptic input similar to that shown in equation 2b. To illustrate the functionality of the K-state (K>3) p-bit network, we compute the Max-4-Cut for the same graph considered in Fig. 2a without the need for graph reduction. The evolution of the 4-Cut solution is presented in Fig. 3.

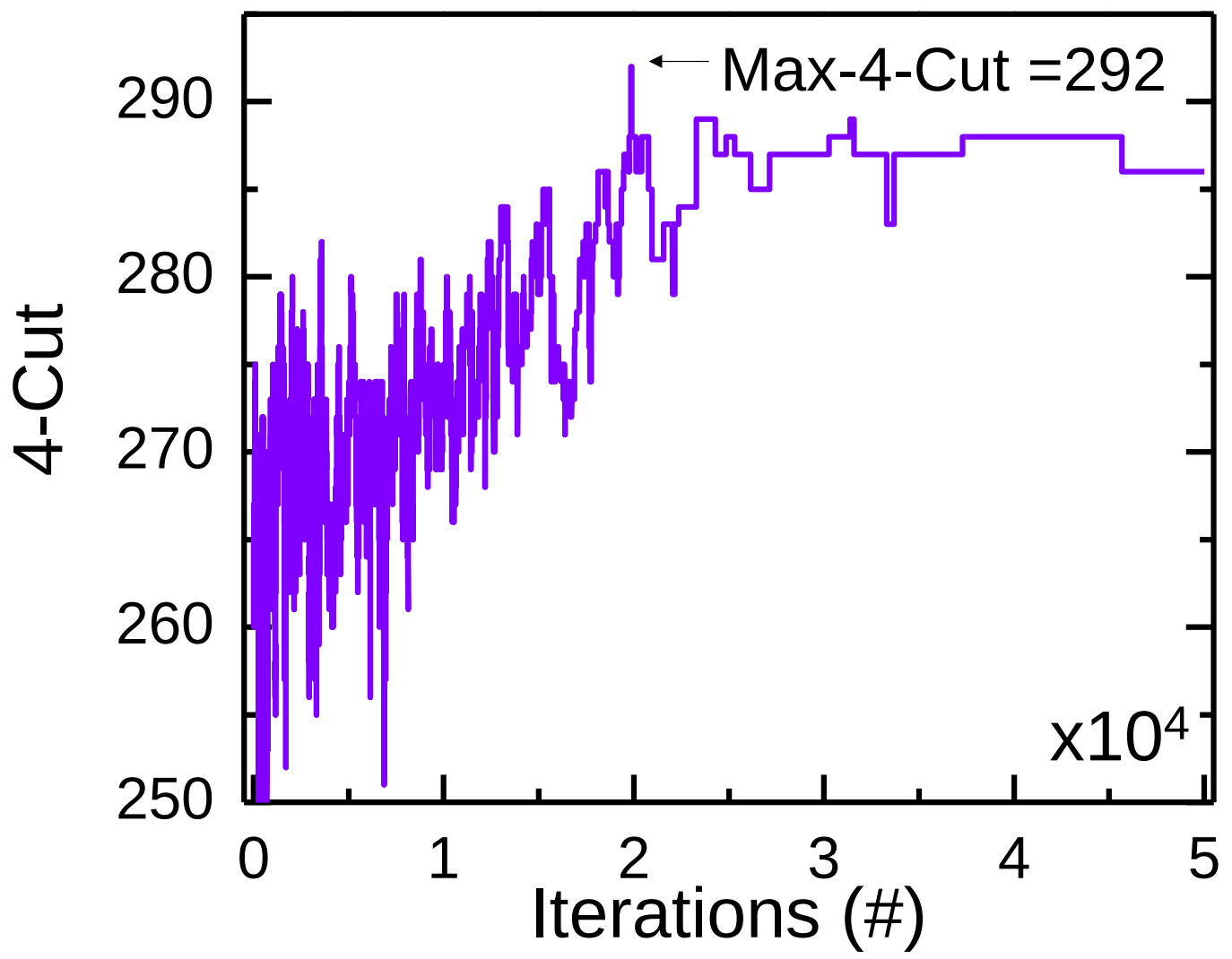

**Fig. 3:** Evolution of 4-Cut with number of iterations, obtained using a 4-state p-bit based solver (β was linearly increased from 0.1 to 0.8 in the simulation).

**Hardware implementation of a K-state p-bit engine:** Equations 2 and 3 help guide the hardware implementation of the K-state p-bit engine. The probabilistic switching behavior in response to the synaptic input that governs whether the state of the p-bit will switch or retain its current state can be realized using the traditional two-state p-bits. In other words, the $f(\phi_\alpha)$ function can be efficiently implemented using the two-state p-bits. However, in addition to this, we also need an additional 'multi-state' p-bit capable of randomly selecting one out of the remaining M states to transition into (if the synaptic input favors such a transition, $f(\phi_\alpha)$ (<0)). Such a multi-state p-bit effectively implements $g$ in equation (3). Furthermore, we note that only one multi-state p-bit is needed for implementing a K-state p-bit engine (Fig. 4). The multi-state p-bit is designed to select one out of M states with uniform probability and its switching probability does not depend upon the specific synaptic input to a particular node. Hence, it can be shared across all the nodes. Thus, a K-state p-bit engine with N nodes can be implemented with *N* two-state p-bits and a multi-state p-bit that selects one of the M states with uniform switching probability. We note that for K=3, the multi-valued p-bit can effectively be implemented using an additional two-state p-bit programmed to switch with a uniform probability of p=0.5.

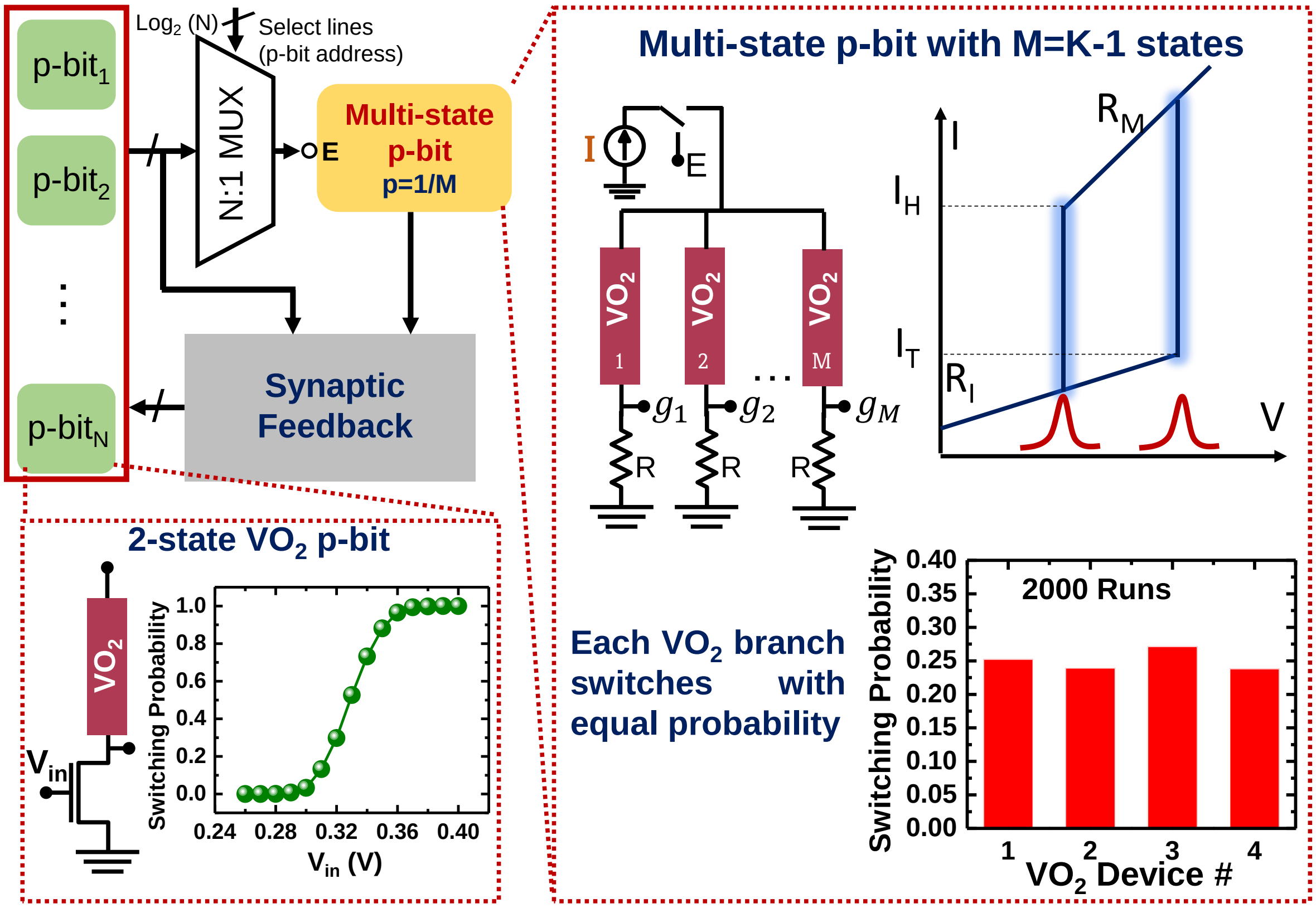

**Fig. 4:** Architecture of a K-state p-bit machine. A K-state p-bit machine can be implemented using N 2-state p-bits and one K-1 (=M)-state p-bit. The structure of a $VO_2$-based 2-state p-bit as well as a $VO_2$-based K-1(=M)-state p-bit is also shown along with their switching characteristics.

The output of a K-state p-bit (say, $\alpha$) can be considered as being composed of two components: (a) the output of the $\alpha^{th}$ two-state p-bit; and (b) the output of the multi-state p-bit (shared among all the bits). These two outputs along with the current state of p-bit can be used to compute the synaptic feedback, and subsequently, update the state using equation 3.

We also note that when computing the synaptic feedback for a node (say $\alpha$), the output of the multi-valued p-bit is only required when the corresponding 2-state p-bit is set to transition out of its original state i.e., $f(\phi_\alpha) < 0$. If $f(\phi_\alpha) > 0$, the node will retain its existing state, thus, eliminating the need for the output of the multi-state p-bit. One approach for implementing this functionality is to use an N:1 MUX, that activates the multi-valued p-bit only when the corresponding p-bit output is 0. However, the MUX design used to control the activation of the multi-state p-bit is optional if multi-valued p-bit is designed to be 'free running'.

Many hardware technologies ranging from MTJs (magnetic tunnel junctions) [16], memristors [17], ferroelectrics, to CMOS [18] have been used to design and implement p-bits. Here, we consider implementing the above K-state p-bit engine using $VO_2$, an insulator-metal phase transition oxide. $VO_2$ exhibits an electronically driven insulator-metal transition (IMT) that is accompanied by abrupt, volatile, and hysteretic resistance switching between the high-resistance insulating state and the low-resistance metallic state. An illustrative figure for the resistance switching is shown in Fig. 4. Moreover, the underlying stochasticity in the nucleation and growth dynamics that govern the abrupt resistance modulation makes the IMT stochastic. Specifically, the trigger voltage ($V_T$: voltage at which the device transitions from the insulating to the metallic state) and hold voltage ($V_H$: voltage at which the device transitions back from the metallic state to the insulating state) exhibit cycle-to-cycle variation, a process that has been extensively studied and characterized in prior works [19]-[23]. Here, we will exploit the variation in $V_T$ to realize the $VO_2$-based p-bit. Empirically, the observed variation in $V_H$ is much smaller than the variation in $V_T$, and therefore, variation in $V_H$ has not been considered here.

Similar to the original MTJ-based two-state p-bit design, the $VO_2$-based p-bit can be implemented using a two-terminal $VO_2$ device in series with a transistor. While the operation of the $VO_2$ p-bit has been detailed in Appendix I, the underlying idea behind the stochastic behavior relies on the probabilistic intersection of the transistor load line with the stable (insulating state) versus the unstable portion of the curve (associated with the transition between the insulating and the metallic state) owing to the cycle-to-cycle variation in the threshold voltage. The latter results in the device switching to the metallic state while in the former case, the device remains in the insulating state. Thus, the probability can be expressed in terms of the probability of switching to the metallic state.

For implementing an M state p-bit, the design entails the use of M parallel branches of the $VO_2$ device, each with a resistive element as shown in Fig. 4; we note that while a resistor has been shown in the figure for simplicity, transistors, biased under the same conditions, could also be used instead. The parallel $VO_2$ branches are driven by a current source whose value is appropriately engineered such that one and only one $VO_2$ branch can be sustained in the metallic state at any given time, while all other branches remain in the insulating state. The constraints on the current, sourced by the current source, as well as the parameters considered for the $VO_2$ in the simulation are described in Appendix II. Owing to the natural stochasticity in the threshold voltage, the device selects one of the $VO_2$ branches with equal probability.

## III. CONCLUSION

In summary, we have developed the underlying framework for realizing a p-bit engine with K-states (K>2). Our approach reduces the computational overhead traditionally associated with a given problem compatible with the two-state behavior of traditional p-bit engines. We also showcase a pathway to implementing a multi-valued p-bit, a key component of the K-state p-bit engine. Thus, our work contributes to enhancing the computational capabilities of p-bit-based computing platforms for solving computationally challenging problems.

## ACKNOWLEDGMENT

This material is based upon work supported by the National Science Foundation under grant no. 2328961 and is supported in part by funds from federal agency and industry partners as specified in the Future of Semiconductors (FuSe) program. The authors wish to thank Professor Supriyo Datta (Purdue University) for valuable comments and feedback on the manuscript.

## COMPETING INTERESTS

The authors declare no competing interests.

## DATA AVAILABILITY

The data that support the findings of this study are available from the corresponding author upon reasonable request.

## AUTHOR CONTRIBUTIONS

Mohammad Khairul Bashar: Software (equal), Investigation (equal), Formal Analysis (equal), Writing – Original Draft (supporting). Abir Hasan: Software (equal), Investigation (equal), Writing – Original Draft (supporting). Nikhil Shukla: Conceptualization (lead), Formal Analysis (equal), Supervision (lead), Project administration (lead), Software (equal), Investigation (equal), Writing – Original Draft (lead).

## Appendix I: Operation of a two-state $VO_2$ p-bit with synaptic input

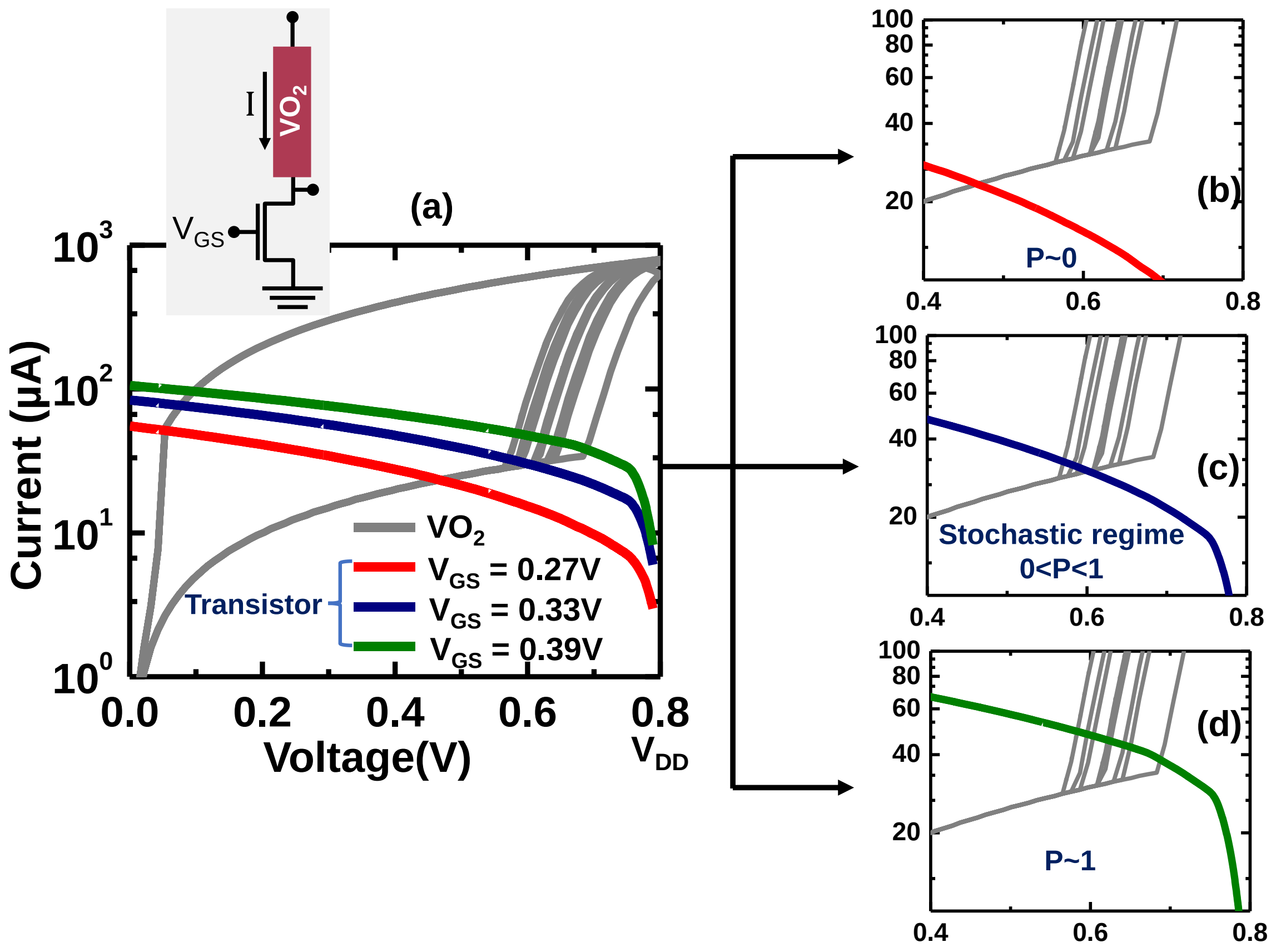

**Fig. A1:** Operation of the $VO_2$ based 2-state p-bit. (a) I-V characteristics of the $VO_2$ device with 3 transistor load lines. (b), (c), and (d) show the transistor load lines for p=0 (no transition), 0<p<1 (stochastic transition), and p=1 (transition), respectively.

Fig. A1 illustrates the operating principle of the proposed $VO_2$-based two-state p-bit using the current versus voltage characteristics of a $VO_2$ device and the series transistor which effectively acts as a 'load line' on the $VO_2$ curve. When the load line intersects the stable insulating state (Fig. A1 (b)) well below $V_T$, no transition is observed. Subsequently, as the gate voltage (synaptic input) is increased and the transistor load line nears the threshold voltage (current) where stochastic cycle-to-cycle variations become significant, the device exhibits stochastic switching behavior (Fig. A1 (c)). Finally, increasing the gate voltage further now results in the load line intersecting with the unstable section of the $VO_2$ I-V curve where the $VO_2$ transitions to the metallic state. This induces a transition to a metallic state (Fig. A1 (d)). The transistor in the p-bit implementation was simulated using the 14nm PTM model. All default parameters, except for PDIBL1=12.3 which effectively lowers the output resistance, were used.

## Appendix II: Constraints on the current source in a $VO_2$-based p-bit

Here, we elucidate the constraints on the current source design used for driving a multi-state p-bit with M states. The constraints on the current can be described by:

$$M\, I_T < I < \frac{(M-1)(R_M+R)+(R_I+R)}{(R_M+R)} I_T \qquad \text{(A1)}$$

The lower bound $M\, I_T < I$ on the current arises from the condition that the current must be larger than the nominal trigger currents to induce IMT in at least one of the $VO_2$ branches. The upper bound on the current source $I < \frac{(M-1)(R_M+R)+(R_I+R)}{(R_M+R)} I_T$, arises from the condition that the current must not be large enough that it induces IMT in more than one $VO_2$ branch. If only one $VO_2$ is in the metallic state, the current through each of the other branches must be smaller than $I_T$. Hence, $\frac{1}{(M-1)} \frac{(R_M+R)}{(R_M+R)+\frac{(R_I+R)}{(M-1)}} I < I_T$, which yields that $I < \frac{(M-1)(R_M+R)+(R_I+R)}{(R_M+R)} I_T$.

The characteristic properties of the $VO_2$ used in the simulation are listed below:

| **$VO_2$ Parameter** | **Value** |
|---|---|
| $R_I$ (Insulating State Resistance) | 20 KΩ |
| $R_M$ (Metallic State Resistance) | 1 KΩ |
| Nominal $I_T$ (Trigger current) | 30 μA |
| Nominal $I_H$ (Hold current) | 50 μA |

The variation in $I_T$ for the $VO_2$ in the 4 branches over 2,000 cycles is shown in Fig. A2.

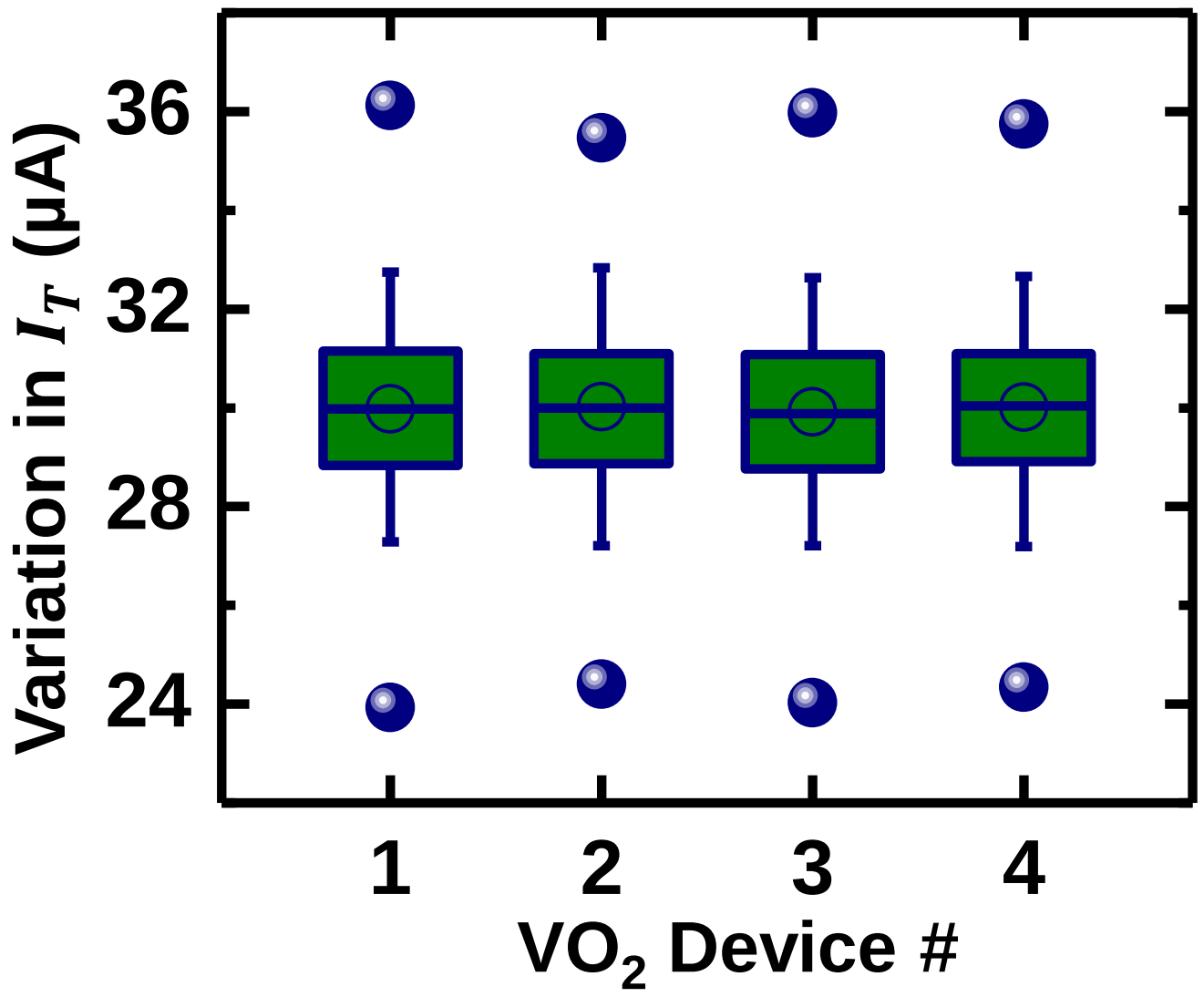

**Fig. A2:** Distribution of $I_T$ for the $VO_2$ devices used to simulate the four-state p-bit shown in Fig. 4 in the main text.

The value of the series resistance in the 4-state $VO_2$-based p-bit is 2 KΩ.